\documentclass[letterpaper, 10 pt, conference]{ieeeconf}  % Comment this line out if you need a4paper

\IEEEoverridecommandlockouts                              % This command is only needed if 
\usepackage{amsmath,epsfig}
\usepackage{amsfonts,amssymb}
\newcommand{\ba}{\left( \begin{array}}
\newcommand{\ea}{\end{array} \right)}
\newcommand{\bq}{\begin{eqnarray*}}
\newcommand{\eq}{\end{eqnarray*}}
\newcommand{\bqn}{\begin{eqnarray}}
\newcommand{\eqn}{\end{eqnarray}}

\usepackage{hyperref}
\hypersetup{
    colorlinks=true,%
    citecolor=blue,%
    filecolor=blue,%
    linkcolor=red,%
    urlcolor=black}

\title{\LARGE \bf Topological Time Frequency Analysis of Functional Brain Signals
}

\author{Moo K. Chung$^1$, Aaron F. Struck$^2$\\
$^1$Department of Biostatistics and Medical Informatics,\\ 
$^2$Department of Neurology\\
University of Wisconsin-Madison
\thanks{*The correspondence should be sent to {\tt mkchung@wisc.edu}. This work was supported by NIH grants EB028753, MH133614 and NSF grant MDS-201077.}}

\begin{document}

\maketitle
\thispagestyle{empty}
\pagestyle{empty}

%%%%%%%%%%%%%%%%%%%%%%%%%%%%%%%%%%%%%%%%%%%%%%%%%%%%%%%%%%%%%%%%%%

\begin{abstract}
We present a novel topological framework for analyzing functional brain signals using time-frequency analysis. By integrating persistent homology with time-frequency representations, we capture multi-scale topological features that characterize the dynamic behavior of brain activity. This approach identifies 0D (connected components) and 1D (loops) topological structures in the signal's time-frequency domain, enabling robust extraction of features invariant to noise and temporal misalignments. The proposed method is demonstrated on resting-state functional magnetic resonance imaging (fMRI) data, showcasing its ability to discern critical topological patterns and provide insights into functional connectivity. This topological approach opens new avenues for analyzing complex brain signals, offering potential applications in neuroscience and clinical diagnostics.
\end{abstract}

\section{Introduction}

Functional brain signals, such as those obtained from electroencephalography (EEG) and functional magnetic resonance imaging (fMRI), exhibit complex temporal and spatial dynamics \cite{chung.2023.NI,lee.2012.tmi,sporns.2003}. Traditional time-frequency analysis methods, such as wavelet transforms and short-time Fourier transforms, have proven effective in capturing signal characteristics over time and frequency domains  \cite{durak.2003}. However, these methods often fall short in robustly identifying topological features, which are crucial for understanding the underlying connectivity and dynamics of brain networks.

Persistent homology, a powerful tool from topological data analysis (TDA), has emerged as an effective approach for studying multi-scale geometric and topological properties of data \cite{chung.2023.NI,edelsbrunner.2000,ghrist.2008}. By identifying and characterizing features such as connected components (0D) and loops (1D) across varying spatial or temporal scales, persistent homology provides a robust and noise-resistant framework for analyzing complex datasets \cite{chung.2023.NI,chung.2019.NN}. Despite its potential, its application to time-frequency analysis of functional brain signals remains underexplored.

In this study, we propose a novel framework that integrates persistent homology with time-frequency analysis to characterize the topological dynamics of functional brain signals. By leveraging this hybrid approach, we extract invariant and interpretable features that capture critical patterns in the brain's dynamic functional connectivity. We demonstrate the effectiveness of our method on resting-state fMRI data, showcasing its ability to discern meaningful topological features across multiple time-frequency scales.
The proposed provides new insights into functional brain connectivity and aims to bridge the gap between topological data analysis and time-frequency analysis, paving the way for innovative applications in neuroscience and clinical diagnostics.

\section{Materials and Methods}

\subsection{Description of Data and Protocol}

In this study, we analyzed a subset of resting-state fMRI data from 400 participants in the Human Connectome Project (HCP) \cite{vanessen.2013}. The age distribution is between 22 and 36 years (average 29.05 $\pm$ 3.36 years). Each fMRI scan lasted approximately 15 minutes, during which participants remained at rest with their eyes open, maintaining a relaxed fixation on a projected bright cross-hair against a dark background. The data were acquired using a customized Siemens 3T Connectome Skyra scanner with a gradient-echo planar imaging (EPI) sequence. Acquisition parameters included a multiband factor of 8, repetition time (TR) of $720$ ms, echo time (TE) of $33.1$ ms, flip angle of $52^\circ$, matrix dimensions of $104 \times 90$ (RO $\times$ PE), 72 slices, and $2$ mm isotropic voxels, with a total of 1200 time points per scan.

MRI preprocessing followed the HCP minimal preprocessing pipelines \cite{glasser.2013}, with additional standard procedures such as motion correction \cite{cox.1996}. A detailed account of our preprocessing pipeline is provided in our previous study \cite{huang.2020.NM}. After preprocessing, the resting-state functional time series was obtained in a volumetric space of $91 \times 109 \times 91$ with $2$ mm isotropic voxels over 1200 time points. Brain parcellation was performed using the Automated Anatomical Labeling (AAL) atlas, segmenting the brain into 116 distinct anatomical regions \cite{tzourio.2002}. The fMRI signal within each parcellated region was averaged across voxels, yielding 116 time series with 1200 time points for each subject.

\subsection{Methods}

\subsubsection{Topological Decomposition of Networks}
For a given graph, we can construct the graph filtration, which is a sequence of nested subgraphs obtained by varying edge weights \cite{lee.2011.MICCAI,chung.2024.NN}. These subgraphs reflect the network's connectivity at different scales. Consider a weighted graph \(\mathcal{X} = (V, w)\), where \(V = \{1, 2, \cdots, p\}\) is the set of nodes corresponding to different regions of the brain, and \(w = (w_{ij})\) represents the edge weights between nodes \(i\) and \(j\). We construct a series of binary networks \(\mathcal{X}_\epsilon = (V, w_\epsilon)\) for different thresholds \(\epsilon\). The binary edge weights \(w_{\epsilon,ij}\) are defined as:

\[ 
w_{\epsilon,ij} = 
\begin{cases} 
1 & \mbox{if } w_{ij} > \epsilon; \\
0 & \mbox{otherwise}. 
\end{cases} 
\]
This process results in a sequence of nested graphs as the threshold \(\epsilon\) is varied, capturing the multiscale structure of the network. For a set of sorted edge weights 
$w_{(1)} < w_{(2)} < \cdots < w_{(q)}$,
the \textit{graph filtration} is given by \cite{lee.2012.tmi,lee.2011.MICCAI}
\[ 
\mathcal{X}_{w_{(1)}} \supset \mathcal{X}_{w_{(2)}} \supset \cdots \supset \mathcal{X}_{w_{(q)}}. 
\]
The graph filtration captures the hierarchical organization of the network's connectivity. The filtration illustrates a spectrum of networks, ranging from the densest, which represents the fully connected network, to the sparsest, which retains only the most significant connections. 

In graph filtrations, 0D topological features, or connected components, emerge when an edge is deleted that disconnect two previously connected nodes at a filtration value \( b_{(i)} \). The set of these birth values forms the 0D persistence barcode \( B = \{ b_{(i)} \} \), with each \( b_{(i)} \) indicating the birth of a new connected component.  During the filtration, all the edges corresponding to birth values form the maximum spanning tree (MST) of the graph  $\mathcal{X}$
\cite{anand.2023.TMI,dakurah.2022}. Conversely, 1D topological features, or loops, are considered to have been born at \(-\infty\) in graph filtrations and die when an edge is added that completes the cycle. This death is captured at the filtration value corresponding to the weight of the edge causing the loop's termination. Accordingly, the 1D persistence barcode consists of points \( D = \{ d_{(i)} \} \), where each \( d_{(i)} \) marks the end of a loop. During the graph filtration, we observe a unique decomposition of edge weights into disjoint sets corresponding to the births of connected components and the deaths of loops, respectively \cite{chung.2023.NI,song.2023}.

\begin{figure}[t]
\begin{center}
\includegraphics[width=1\linewidth]{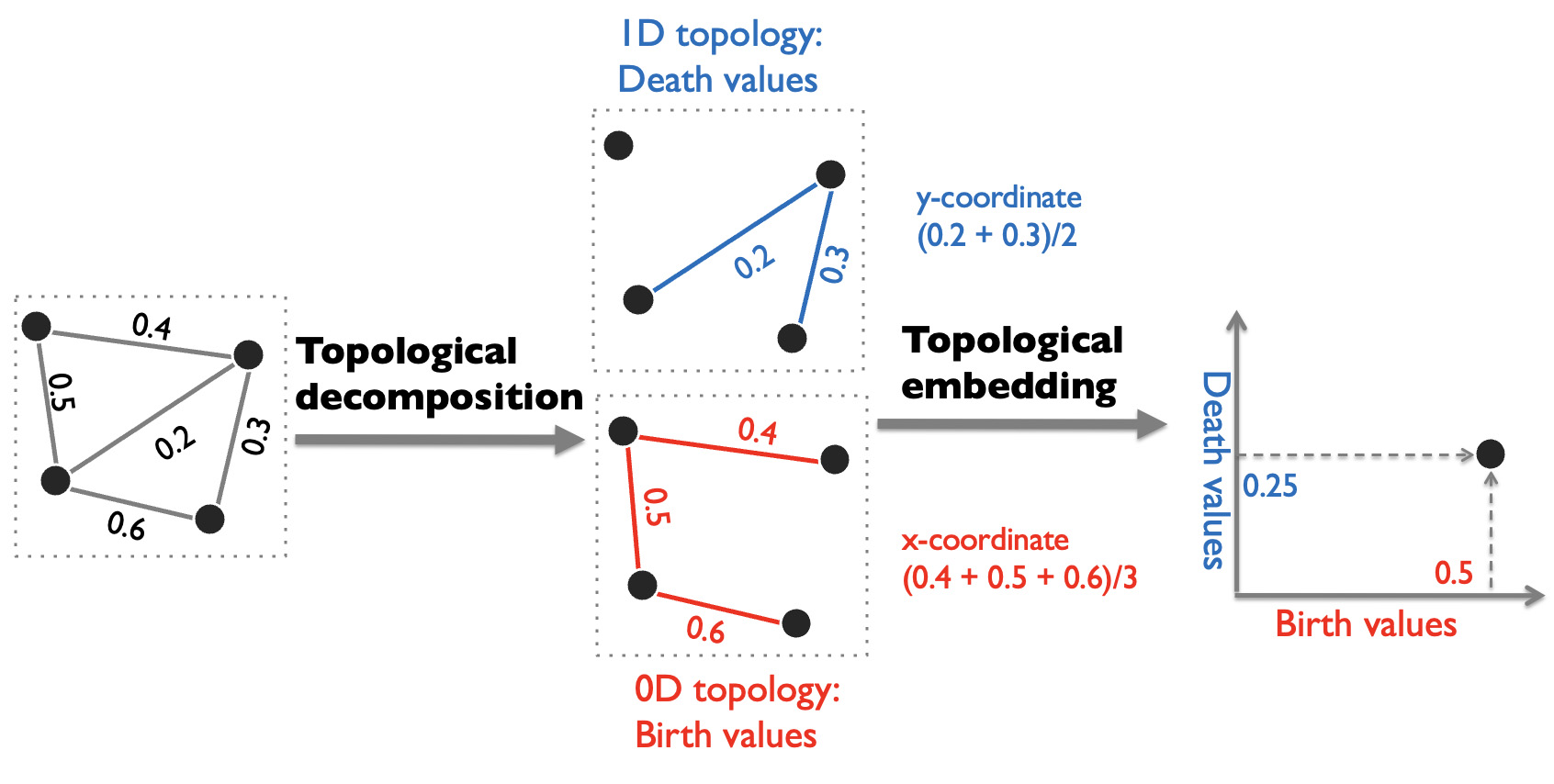}
\caption{The overall schematic of topological decomposition  followed by topological embedding. The birth values correspond to the maximum spanning tree (MST) of the underlying graph while the death values correspond to none MST edges. The process simplifies the weighted weighted network into a topological summary represented as a 2D point, where $x$-axis in TPD represents the average birth values, while the $y$-axis represents the average death values of the network.}
\label{fig:schematic}
\end{center}
\end{figure}

For a graph \(\mathcal{X} = (V, w)\) with an edge weight set \( W  = \{ w_{(1)}, \cdots, w_{(q)} \} \), there exists a unique decomposition:
$$W = B \cup D, \quad B \cap D = \emptyset,
$$
where the birth set \( B = \{ b_{(1)}, b_{(2)}, \cdots, b_{(q_0)} \} \) is the collection of 0D sorted birth values, and the death set \( D = \{ d_{(1)}, d_{(2)}, \cdots, d_{(q_1)} \} \) is the collection of 1D sorted death values, with \( q_0 = p-1 \) and \( q_1 = \frac{(p-1)(p-2)}{2} \). 
Fig. \ref{fig:schematic} provides the schematic of birth-death decomposition of a weighted graph.

\subsubsection{Topological Time Frequency Analysis}
\label{sec:TFA}

Traditional approaches to analyzing time-varying rs-fMRI connectivity often rely on comparing connectivity matrices at identical time points across subjects \cite{cohen.1995}. However, this is problematic due to the inherent asynchrony of spontaneous neural dynamics particularly in the resting state. Inter-individual variability renders such frame-wise comparisons both statistically and biologically unreliable. To address the challenge, we propose a \emph{Topological Time Frequency Analysis} framework that bypasses the need for temporal alignment by focusing on the evolution of topological features—such as connected components and cycles—within each subject’s dynamic brain network \cite{chung.2024.foundations}. These features are summarized over time using persistent homology, and their evolution is projected into the frequency domain using the Fourier transform. The resulting topological spectrogram captures periodic or rhythmic modulations in network topology and provides a synchronization-invariant, multiscale view of brain dynamics.

\begin{figure}[b]
\begin{center}
\includegraphics[width=1\linewidth]{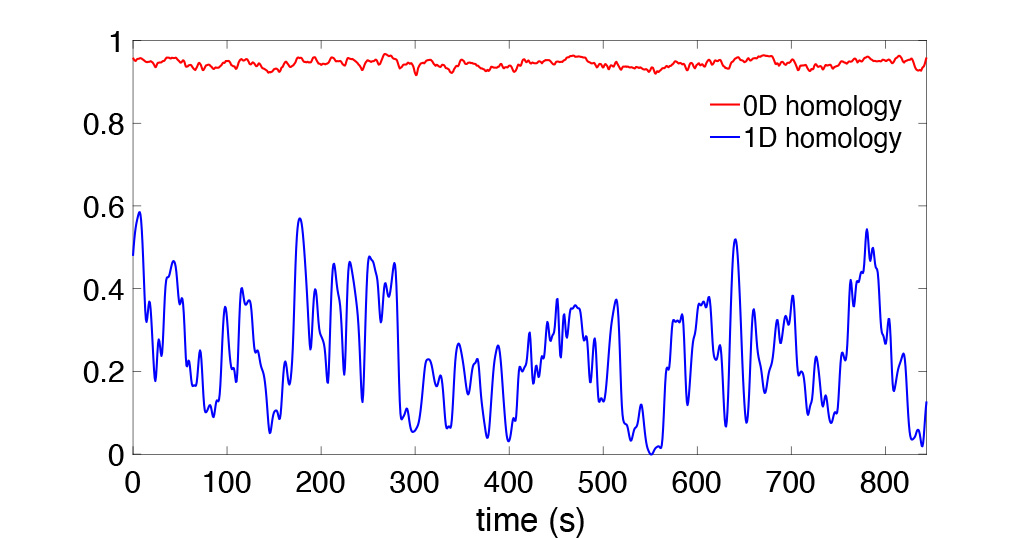}
\caption{Average cumulative birth value $b(t)$ (red) and death value $d(t)$ (blue) for a representative subject. 0D homology, which is determined by the maximum spanning tree, remains very stable. 
In contrast, 1D homology fluctuates rapidly relative to 0D homology, reflecting cyclical changes in the rs-fMRI network over time.
Their spectrograms will reveal that 0D homology remains stationary, whereas 1D homology is non-stationary.}
\label{fig:cb}
\end{center}
\end{figure}

Since we can decompose topology into 0D and 1D orthogonally, we can analyze 0D and 1D topology separately. Thus, we perform {\em topological embedding} into a 2D plane (Fig. \ref{fig:schematic}). In the topological embedding, the $x$-axis represents the spread with respect to the 0D topology, while the $y$-axis represents the spread with respect to the 1D topology \cite{chung.2023.NI}. For a time-varying network \(\mathcal{X}_t\), the embedding $x$- and $y$-coordinates at time $t$ are given by the average cumulative birth and death values:
$$
b(t) = \frac{1}{q_0} \sum_{i=1}^{q_0} b_{(i)}^t, \quad d(t) = \frac{1}{q_1} \sum_{i=1}^{q_1} d_{(i)}^t, \label{eq:y_t}
$$
where $b_{(i)}^t$ and $d_{(i)}^t$ are the $i$-th smallest birth and death values of the network at time $t$.
Fig. \ref{fig:cb} displays a representative average cumulative birth values $b(t)$ and death values $d(t)$.  This provides the topological summary of how functional brain networks changes over time.

\begin{figure}[t]
\begin{center}
\includegraphics[width=1\linewidth]{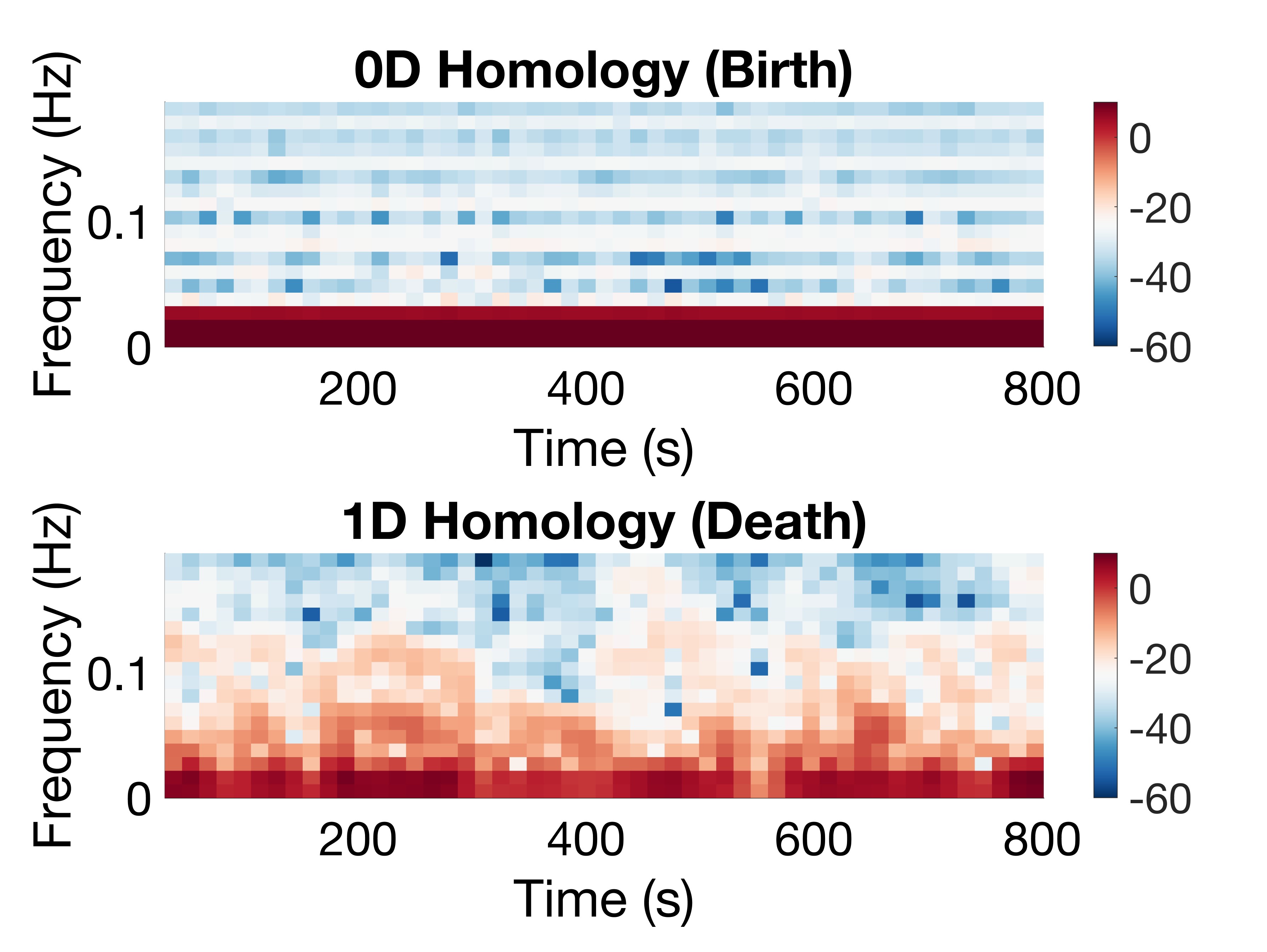}
\caption{The topological spectrogram for a representative subject. The power spectral
density (PSD) levels indicate the intensity of topological signals across the frequency spectrum. The consistently high PSD in the low-frequency range (0-0.03 Hz) for 0D homology suggests a persistent presence of a tree-like backbone (maximum spanning tree), which remains stable throughout the observation period. In contrast, the 1D homology exhibits more variable signals, corresponding to the dynamic formation and dissolution of cycles as nodes within the tree connect or disconnect.
This variability underscores the transient nature of higher-dimensional topological structures in the brain networks.
}
\label{fig:TFA}
\end{center}
\end{figure}

Then we perform the Short-Time Fourier Transform (STFT) for 0D and 1D topology:
\bq
B(\tau, \omega) &=& \int b(t) w(t-\tau) e^{-j \omega t} dt, \\
D(\tau, \omega) &=& \int d(t) w(t-\tau) e^{-j \omega t} dt,
\eq
where $w(t)$ is the window function, $\tau$ is the time shift, and $\omega$ is the angular frequency. The power spectral density (PSD) in decibels (dB) is:
\bq
P_{B}(\tau, \omega) &=& 20 \log_{10} B(\tau, \omega), \\
P_{D}(\tau, \omega) &=& 20 \log_{10} D(\tau, \omega).
\eq
Fig. \ref{fig:TFA} display the topological spectrogram for 0D and 1D topology for a single subject. The 0D topology (MST) is stable across time, while the 1D topology fluctuates, connecting and disconnecting edges around the MST, forming or destroying cycles. The patterns are consistently observed in all 400 subjects in the study.

\subsubsection{Validation on Simulations}
Since real brain network data lack the mathematical ground truth, we conducted extensive validations through statistical simulations. We validated the performance of the proposed topological features (births and death) against existing methods ($k$-means and hierarchical clustering) in a clustering task. 

\begin{figure}[t]
\centering
\includegraphics[width=1\linewidth]{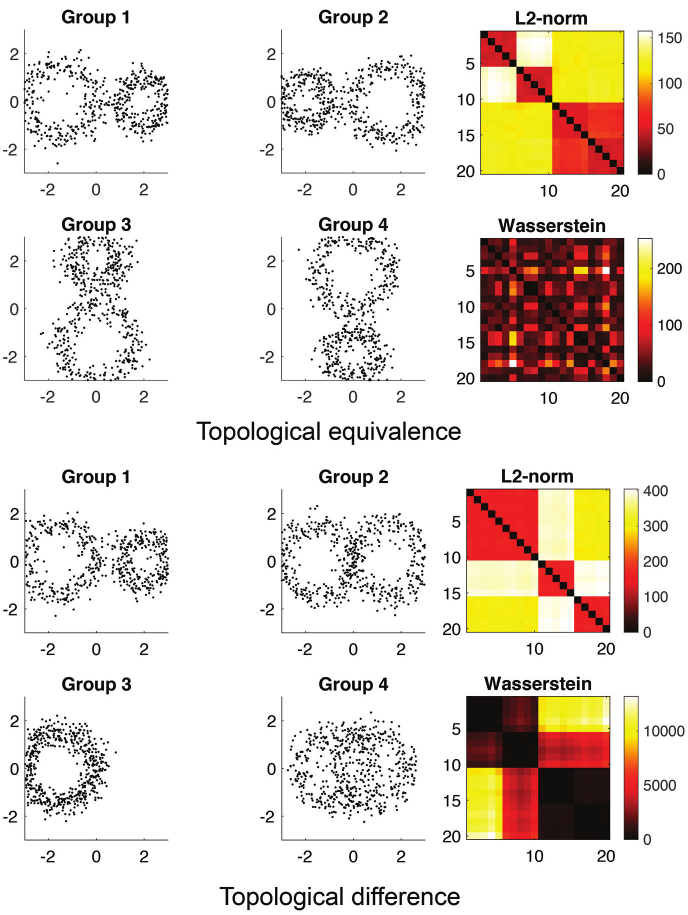}
\caption{Top: simulation study on  topological equivalence. The correct clustering method should {\em not} be able to cluster them because they are all topologically equivalent. The pairwise Euclidean distance ($L_2$-norm) is used in $k$-means and hierarchical clustering. The Wasserstein distance is used in topological clustering. Bottom: simulation study on  topological difference. The correct clustering method should be able to cluster them because they are all topologically different.}
\label{fig:simulationnodiff}
\end{figure}

We generated four groups of circular patterns, both topologically identical (Fig. \ref{fig:simulationnodiff}-top) and topologically different (Fig. \ref{fig:simulationnodiff}-bottom). Each circle was uniformly sampled with 60 nodes, and independent Gaussian noise $N(0, 0.3^2)$ was added to each coordinate. Five random networks were generated per pattern. Fig. \ref{fig:simulationnodiff} displays the superimposed nodes from 20 networks. The Euclidean distance ($L_2$-norm) between nodes was used to construct connectivity matrices for $k$-means and hierarchical clustering. For topological features, we used the Wasserstein distance, which is the most widely used distance for topological features in persistent homology \cite{song.2023}. For sorted birth and death values $b_{(i)}^k, d_{(i)}^k$  for the $k$-th network $\mathcal{X}^k$, We used the 2-Wasserstein distance is given by 
$$d(\mathcal{X}^1, \mathcal{X}^2)^2 = \sum_{i=1}^{q_0}  \big( b_{(i)}^1 - b_{(i)}^2 \big)^2 + \sum_{i=1}^{q_1} \big( d_{(i)}^1 - d_{(i)}^2 \big)^2.$$

In the experiment given in Fig. \ref{fig:simulationnodiff} (top), we assessed the occurrence of {\it false positives} in patterns lacking topological differences. All groups, derived through rotations, are topologically identical. Therefore, any detected differences are false positives. While $k$-means clustering showed an accuracy of $0.90 \pm 0.15$ and hierarchical clustering achieved perfect accuracy of 1.00, both methods reported significant false positives, \underline{incorrectly} clustering each pattern as distinct clusters.  In contrast, topological clustering, with an accuracy of $0.53 \pm 0.08$, showed 37\% fewer false positives in the absence of topological differences. Thus, $k$-means clustering and hierarchical clustering are ill-suited for learning tasks in the presence of topologically similar objects.

Fig. \ref{fig:simulationnodiff} (bottom) illustrates our test for {\it false negatives}, involving distinct topological patterns with varying numbers of cycles and distinct topologies. In this scenario, topological differences should be detected. $k$-means clustering recorded an accuracy of $0.83 \pm 0.16$, while hierarchical clustering again reported perfect accuracy of 1.00. Topological clustering also attained a high accuracy of $0.98 \pm 0.09$.

Since any topologically distinct object should exhibit local geometric differences, testing for topological differences is straightforward for most methods. However, traditional clustering methods are prone to a significant number of false positives, making them less suitable for topological learning tasks. The proposed topological distance-based approach demonstrates superior performance, particularly in reducing false positives by 47-50\% compared to traditional methods using $k$-means and hierarchical clustering.

\section{Results and Discussion}

%\begin{figure}[t]
%\begin{center}
%\includegraphics[width=1\linewidth]{TE.jpg}
%\caption{\small Topological embedding illustrating the temporal evolution of a functional brain network for two representative subjects (colored red and white). 
%The $x$-axis represents the normalized cumulative birth values $b(t)$, while the $y$-axis represents the normalized cumulative death values $d(t)$.
%The background shows the binned 2D histogram of all 400 subjects. Although there is significant individual variability, the data generally clusters around a unimodal peak at approximately (0.94, 0.2).}
%\label{fig:TPD}
%\end{center}
%\end{figure}

\begin{figure}[t]
\begin{center}
\includegraphics[width=1\linewidth]{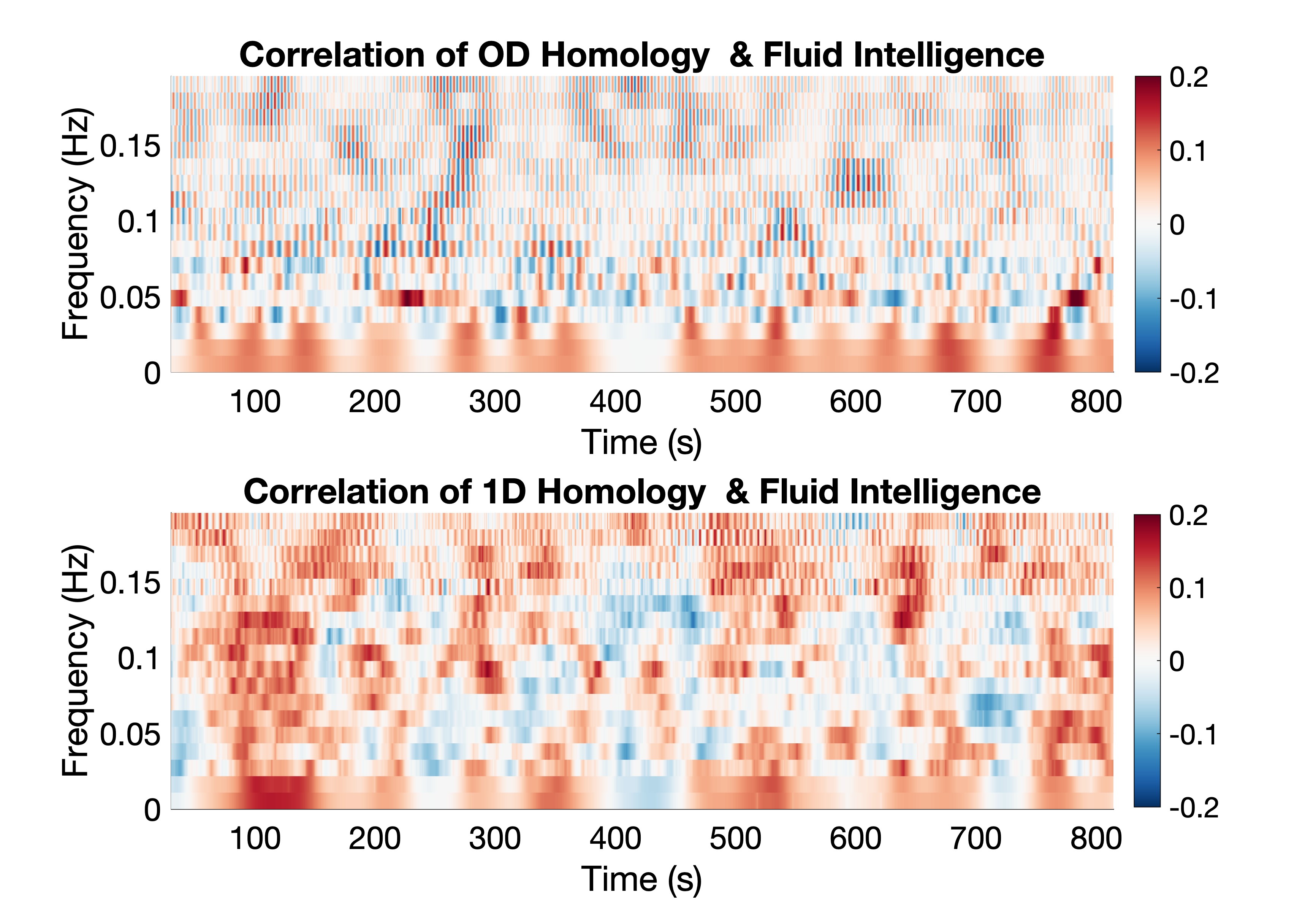}
\caption{Correlation between the spectrogram of 0D and 1D topologies and fluid intelligence in 400 subjects. A statistically significant but weak positive correlation was observed in the 0-0.02 Hz range for the 0D topology, indicating that very slow, more stable oscillatory patterns in 0D topology are associated with fluid intelligence. In contrast, 1D topology correlates with fluid intelligence across a broader frequency range (0-0.2 Hz), suggesting that dynamic, recurrent network interactions and higher-frequency fluctuations contribute to cognitive processing. The broader frequency involvement of 1D topology may reflect the role of cyclic neural activity in adaptive problem-solving and executive function.}
\label{fig:fluid}
\end{center}
\end{figure}

\begin{figure}[t]
\begin{center}
\includegraphics[width=1\linewidth]{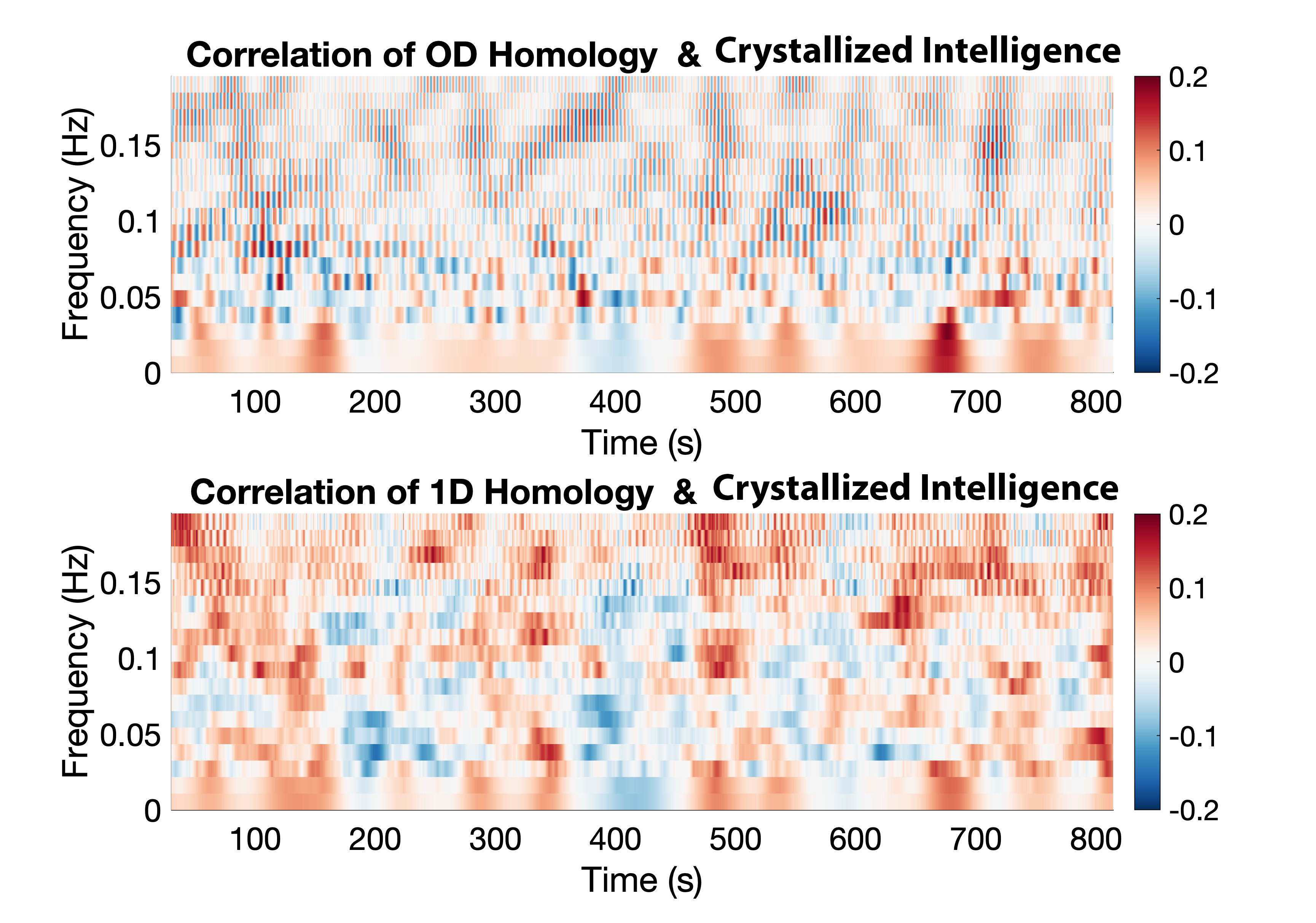}
\caption{Correlation between the spectrogram of 0D and 1D topologies and crystallized intelligence in 374 subjects. A weak positive correlation was observed in the 0-0.02 Hz range for the 0D topology, suggesting that slow, stable oscillatory patterns may contribute to the retention and retrieval of accumulated knowledge. In contrast, 1D topology correlates with crystallized intelligence across a broader frequency range (0-0.2 Hz), indicating that more dynamic and cyclic neural interactions play a role in integrating and accessing stored information.}
\label{fig:crystalized}
\end{center}
\end{figure}

Pearson correlations were computed over the sliding window of size 28 TRs (20.16 seconds) to produce dynamically changing correlation brain netework used in this study \cite{shirer.2012,allen.2014,leonardi.2015}.  Each subject had 1173 time-varying correlation matrices of size 379 x 379 for analysis. We then performed the topological embedding on 400 subjects using time varying correlation matrices. %Fig. \ref{fig:TPD} displays the topological embedding of two representative subjects, illustrating how dynamically changing rs-fMRI connectivity networks are represented as evolving curves in the 2D plane.
Subsequently, we computed the topological spectrogram for each subject. For the Short-Time Fourier Transform (STFT), we used a window size equivalent to 10\% of the total signal duration. The sampling frequency was set to 1.389 Hz, corresponding to the repetition time (TR) of 0.72 seconds in the HCP dataset. The overlap between consecutive windows was maximized to window size minus one sample, effectively allowing smooth transitions between segments. The Fourier Transform was computed using a Fast Fourier Transform (FFT) length set to the next power of two greater than the window size. These parameter choices allow for a detailed yet stable representation of time-frequency dynamics in the topological spectrogram. Fig. \ref{fig:TFA} displays the spectrogram for one representative subject. The pattern is similar for all other subjects. We observed consistent activity in the frequency bands between 0-0.03Hz for 0D topology, which suggests a common, stable pattern of MST, indicative of a fundamental underlying brain network backbone that persists across individuals and time. In contrast, for 1D topology, observed frequencies between 0-0.03Hz reflect the transient topological changes associated with the birth and death of cycles.

We further assessed the relevance of topological metrics in capturing cognitive differences by correlating the topological spectrogram with fluid and crystallized intelligence across 0D and 1D homology in 374 subjects \cite{irby.2013,johnson.2008}. Fluid intelligence refers to the capacity to reason abstractly, solve novel problems, and think abstractly without relying on prior knowledge or experience. It is assessed in the HCP using the Penn Progressive Matrices (PMAT24-A) test, which is a component of the NIH Toolbox Cognition Battery. In contrast, crystallized intelligence reflects accumulated knowledge, learned skills, and the ability to apply past experiences to problem-solving. In the HCP, crystallized intelligence is assessed using the Picture Vocabulary Test, another component of the NIH Toolbox Cognition Battery. 

The threshold for statistical significance is set at a correlation magnitude of \(\pm 0.2\), which corresponds to a one-sided \(p\)-value of 0.0018. The significance level is determined using the \emph{t-test for Pearson correlation}, where the test statistic is computed as
\[
t = r \cdot \sqrt{\frac{n - 2}{1 - r^2}},
\]
with \(t\) following a Student's \(t\)-distribution with \(n-2\) degrees of freedom under the null hypothesis \(H_0: r = 0\) \cite{chung.2007.corr}.

%\begin{figure}[t]
%\begin{center}
%\includegraphics[width=1\linewidth]{spectogram-corr-age.jpg}
%\caption{Correlation of topological spectrogram with age. The spectrogram exhibits a negative correlation with age, indicating reduced neural oscillatory power and less structured brain dynamics in aging. 0D homology shows significant negative correlation in the low-frequency range (0-0.02 Hz), suggesting declining global connectivity stability. In contrast, 1D homology exhibits broader negative correlations across 0-0.17 Hz, reflecting widespread reductions in recurrent network interactions. Correlations exceeding $\pm$0.2 are statistically significant at $p = 0.0018$.}
%\label{fig:corrstrage}
%\end{center}
%\end{figure}

In general, fluid intelligence exhibits stronger correlations with the spectrogram than crystallized intelligence (Fig. \ref{fig:fluid} and \ref{fig:crystalized}), suggesting that dynamic brain activity is more relevant for problem-solving and reasoning abilities than for accumulated knowledge. While 0D homology shows correlations with both types of intelligence, primarily in the 0-0.02 Hz range, 1D homology demonstrates broader correlations across the entire 0-0.2 Hz frequency range. This pattern suggests that low-frequency fluctuations in network connectivity (captured by 0D homology) may contribute to both types of intelligence, whereas higher-order cyclic interactions (captured by 1D homology) are more strongly linked to fluid intelligence across multiple frequency scales. The broader frequency range of 1D homology correlations may reflect the involvement of recurrent network dynamics and long-range feedback loops, which are crucial for adaptive cognitive processing. This suggests that individuals with higher intelligence exhibit more pronounced or structured cyclic fluctuations in their functional brain signals.

\section{Conclusion}

We introduced a novel topological framework for analyzing functional brain signals by integrating topological data analysis with time-frequency representations. This approach captures multi-scale topological features that characterize the dynamic structure of functional connectivity. By leveraging persistent homology, we extracted 0D (connected components) and 1D (loops) topological features from time-varying brain networks, enabling a more robust and interpretable analysis of brain signal dynamics. 

The proposed method offers a novel perspective on brain network dynamics by emphasizing their topological evolution rather than relying solely on direct signal comparisons. This approach has broad implications for understanding cognitive function, brain disorders, and personalized neuroscience applications. Future work will extend this framework to other cognitive variables and explore its applicability to clinical populations, including epilepsy and Alzheimer's disease.

\bibliography{reference.2025.04.20}
\bibliographystyle{unsrt}

\end{document}